\documentclass[%
 reprint,
 amsmath,amssymb,
 aps,
]{revtex4-2}

\usepackage{graphicx}
\usepackage{dcolumn}
\usepackage{bm}
\usepackage{hyperref}

\hypersetup{colorlinks=true, citecolor=blue, urlcolor=blue, linkcolor=blue}

\begin{document}

\preprint{APS/123-QED}

\title{Effect of electric field on excitons in wide quantum wells}

\author{Shiming Zheng}
\email{zsm210603@gmail.com}
\author{E. S. Khramtsov}%
\author{I. V. Ignatiev}%
\affiliation{Spin Optics Laboratory, St. Petersburg State University, Ulyanovskaya 1, Peterhof, St. Petersburg, 198504, Russia}%

\date{\today}

\begin{abstract}
A microscopic model of a heterostructure with a quantum well (QW) is proposed to study the exciton behavior in an external electric field. The effect of an electric field ranging from 0 to 6 kV/cm applied to the GaAs/AlGaAs QW structure in the growth direction is studied for several QWs of various widths up to 100 nm. The three-dimensional Schr{\"o}dinger equation (SE) of exciton is numerically solved using the finite difference method. Wave functions and energies for several states of the heavy-hole and light-hole excitons are calculated. Dependencies of the exciton state energy, the binding energy, the radiative broadening, and the static dipole moment on the applied electric fields are determined. The threshold of exciton dissociation for the 100-nm QW is also determined. In addition, we found the electric-field-induced shift of the center of mass of the heavy-hole and light-hole exciton in the QWs. Finally, we have modeled reflection spectra of heterostructures with the GaAs/AlGaAs QWs in the electric field using the calculated energies and radiative broadenings of excitons.
\end{abstract}

\maketitle

\section{\label{sec:level1}Introduction}
\vspace{-2mm}
An exciton in a semiconductor crystal is a quasiparticle consisting of an electron and a hole. The first observation of exciton states was done by E.F. Gross and his team in 1952~\cite{Gross-1952,Gross-PhU1962}. According to basic semiconductor physics, the hole that makes up the exciton can be classified as either a heavy-hole (hh) or a light-hole (lh) in the GaAs-type crystals. Correspondingly, in the quantum-well (QW) structures, there are two types of excitons, those with heavy holes (Xhh) and those with light holes (Xlh)~\cite{Ivchenko2005}. Modern technology has made it easy to create high-quality heterostructures with QWs, allowing for the research of many exciton characteristics in experimental work~\cite{Gibbs-NPh2011,Bataev-PRB2022}. 

Two ways to control an exciton state are applying an electric field to the heterostructure and changing the width of the QW, in which the exciton excites~\cite{Zhu-PRB1988,Khramtsov-PRB2019}. Energies of the exciton states change with the applied electric field, resulting in the Stark effect for the exciton~\cite{Miller-PRL1984}.

There have already been many studies published on the influence of electric fields on excitons in QWs~\cite{Mendez-PRB1982,Miller-PRL1984,Matsuura-PRB1986,Zhu-PRB1988,Fafard-PRB1993,Gorbunov-JETP-Lett2012,Dorow-APL2018,Loginov-PRResearch2020,Chukeev-PRB2024}. They are mainly devoted to the GaAs/Al$_{0.3}$Ga$_{0.7}$As QWs with relatively small widths ($< 20$ nm) comparable with the Bohr radius of the exciton. In the wider QWs, however, the effect of the electric field on the center-of-mass motion of the exciton can be observed \cite{Dorow-APL2018,Loginov-PRResearch2020}. Moreover, due to the finite height of the potential of the barrier layer, the two-dimensional exciton in the QW acquires properties of a three-dimensional object \cite{Belov-Conf-2018}.

Applying a constant electric field of controllable magnitude in practical work can be a challenging problem due to the nature of capacitors, which are composed of heterostructures. Namely, there is some amount of free charge carriers in real heterostructures, which can be moved by the applied voltage, thus partially compensating the electric field. One of the ways to determine the exact value of the applied electric field is to compare the experimental reflection spectra of the heterostructure under study with the calculated one. By this way, Chukeev et al.~\cite{Chukeev-PRB2024} had experimentally studied the effect of electric fields on exciton in a single QW in the heterostructure. In the cited paper, the authors experimentally realized electric fields up to 5 kV/cm in a 30-nm QW. When the electric field $F = 0$, they clearly observed in the reflection spectrum the lowest quantum-confined states of the heavy-hole (Xhh1) and light-hole (Xlh1) excitons only. However, for the non-zero field, the second quantum-confined state (Xhh2) becomes clearly visible. Additionally, the authors tried to investigate the exciton dipole moment arising in the electric field but did not explore the relationship between the exciton dipole moment and the width of the QW.

The goal of this paper is to theoretically study in depth the dependence of the energy, optical coupling, and dipole moment of several exciton states in the relatively wide QWs on the applied electric field. To contrast with narrow QWs studied earlier~\cite{Miller-PRL1984,Fafard-PRB1993,Mendez-PRB1982,Gorbunov-JETP-Lett2012,Matsuura-PRB1986}, the relatively small electric field strength of order of several kV/cm is required to obtain the change of potential energy of a charge carrier of about exciton binding energy, $E_b$, at a distance of about the exciton Bohr radius. For example, the electric field, $F = 6$ kV/cm, gives rise to the energy change, $E = 4.3$ meV, at $a_B \approx 15$ nm, which is the Bohr radius of an exciton in the GaAs crystal. This energy change is close to the $E_b$ in the bulk GaAs. Correspondingly, the QWs of width $L > 2a_B \approx 30$ nm can be considered as intermediate in width between the narrow and wide QWs~\cite{Ivchenko2005}.

A rectangular GaAs/Al$_{0.3}$Ga$_{0.7}$As QW with a finite depth of potential well is considered. The heterostructure under consideration contains a single QW with a width of 30, 50, or 100 nm. The QW with an intermediate width of $L = 30$ nm is considered since the electric field effects in this QW can be described by perturbation theory. In wider QWs, $L = 50$ and $100$ nm, there is a strongly nonlinear behavior of exciton properties. Thus, we can trace the transition from a relatively narrow QW to wide QWs.

Microscopic modeling was conducted to analyze the exciton states under electric fields ranging from 0 to 6 kV/cm. The electric field is supposed to be applied in the direction of the heterostructure growth. We consider several exciton states, Xhh1, Xlh1, and Xhh2, that can be experimentally identified and studied for QWs of this width. For wider QWs, exciton resonances may overlap~\cite{Khramtsov-PRB2019}, which complicates the interpretation of the results.

The exciton states in the QW were described by solving the three-dimensional SE numerically using the fourth-order finite difference method (FDM). The energies and wave functions of the exciton states were obtained for various electric field strengths. The wave functions were then used to calculate the radiative broadening and static dipole moments of the exciton using well-known formulas~\cite{Ivchenko2005}. We describe the calculation principle and corresponding results for the static exciton dipole moment. The dipole moment appears due to the stretching of the exciton in an electric field.

\section{Basic theory}
\vspace{-2mm}
A single electron can be described in the Cartesian coordinate system by three spatial coordinates ($x$, $y$, and $z$). Hereafter it is assumed that the $x$ and $y$ coordinates are directed along the QW layer and the coordinate $z$ is directed along the growth axis of the heterostructure. Because of degeneration of the valence band in the GaAs-type crystals, the hole states are described by the Luttinger Hamiltonian~\cite{Luttinger-PR1955}. Correspondingly, the effective mass of the hole becomes anisotropic and can be expressed via the Luttinger parameters. For instance, and the effective mass of a hole along the $z$ coordinate is $m_{hz}$. The effective mass of a hole in the $xy$ plane is assumed to be isotropic, resulting in the relation $m_{hx}=m_{hy}=m_{hxy}$. The effective mass of an electron is isotropic in all three directions and is equal to $m_e$.

We consider the QWs of finite width when the heavy-hole and light-hole states are split by the quantum confinement effect. In this case, we can neglect the effects of the heavy-hole-light-hole mixing and consider the Xhh and Xlh excitons separately. The analysis carried out in a number of works~\cite{Collins-1987,Andreani-PRB1990,Bataev-PRB2022,Grigoryev-JETP2023} shows that the mixing effects are small compared to the electric-field-induced effects discussed in this paper. Therefore, we consider here only the diagonal part of the Luttinger Hamiltonian~\cite{Khramtsov-JPh2016}. We also do not account for the spin interaction within the exciton. This is not important when the magnetic field is not considered~\cite{Grigorieva-Physics-of-the-Solid-State2023}. 

The time-independent six-dimensional SE can be used to describe the quantum mechanical behavior of each type of exciton, taking into account the Coulomb interaction between the hole and the electron. We don't consider here the small discontinuity of the dielectric constants of the QW and the barriers. This effect is negligible for the relatively wide QWs we consider~\cite{Pokutny-S2007,Belov-Jph2017}. The SE for an exciton reads:
%
\begin{equation}
\label{eq1}
    \begin{aligned}
        &\left[K(x_h,y_h,z_h,x_e,y_e,z_e)-\frac{q^2}{\varepsilon \cdot r}+V_e(z_e)+V_h(z_h)\right]\\
        &\psi(x_h,y_h,z_h,x_e,y_e,z_e)=E \cdot \psi(x_h,y_h,z_h,x_e,y_e,z_e).
    \end{aligned}
\end{equation}

Here, $K$ is the exciton kinetic operator; $\varepsilon$ is the dielectric constant; $r$ is the distance between the electron and the hole; $q$ is the absolute value of the electron charge; $V_e$ is the potential barrier for the electron; $V_h$ is the potential barrier for the hole; $\psi$ is the wave function of the exciton; $E$ is the energy of the exciton; $\hbar$ is reduced Planck's constant; $z_e$ and $z_h$ are the $z$ coordinates for the electron and the hole, respectively.

The operator $K$ that pertains to exciton kinetics energy can be expressed as follows:
%
\begin{equation}
\label{eq2}
\begin{aligned}
&K=
-\frac{\hbar^2}{2m_{e}}  \nabla_e^2
-\frac{\hbar^2}{2m_{hxy}}\nabla_h^2
-\frac{\hbar^2}{2m_{e}}\frac{\partial^2}{\partial z_e^2}
-\\
&\frac{\hbar^2}{2m_{hxy}}\frac{\partial^2}{\partial z_h^2}. 
\end{aligned}
\end{equation}
Here $\nabla^2$ is the Laplace operator in the $xy$ plane.

The electric field is assumed to be applied along the $z$ axis, coinciding with the growth axis of a heterostructure. In the presence of an electric field, the QW potential profile for the electron and the hole is transformed from rectangular to trapezoidal. Therefore, we can express the potential energy for the electron and hole as follows:
%
\begin{eqnarray}
\label{eq3_Ve_Vh}
V_e&=-q F z_e+&
\left\{
  \begin{gathered}
    0,\quad \text{if}\quad |z_e|<L/2\\
    V_{e0},\quad \text{if}\quad |z_e|\geq L/2
  \end{gathered}
\right.
\\
V_h&=q F z_h+&
\left\{
  \begin{gathered}
    0,\quad \text{if}\quad |z_h|<L/2\\
    V_{h0},\quad \text{if}\quad |z_h|\geq L/2
  \end{gathered}
\right. 
\end{eqnarray}
Here $F$ is the electric field strength, and $V_{e0}$ and $V_{h0}$ are, respectively, the potential energies of the electron and the hole outside the QW.

The exciton can freely move along the QW layer. Correspondingly, we can introduce the coordinate of the center-of-mass motion in the $xy$ plane:
%
\begin{equation}
\begin{aligned}
    X= \frac{m_h \cdot x_{hxy}+m_e \cdot x_e}{m_{hxy}+m_e};Y=\frac{m_h \cdot y_{hxy}+ m_e \cdot y_e}{m_{hxy}+m_e}.
\end{aligned}
\end{equation}
Using these variables, the exciton wave function can be factorized:
%
\begin{equation}
\label{eq6}
\begin{aligned}
    \psi(x_h,y_h,z_h,x_e,y_e,z_e)=\psi(z_e,z_h,\rho,\theta)e^{iK_XX}e^{iK_YY},
\end{aligned}
\end{equation}
where $e^{iK_XX}$ and $e^{iK_YY}$ are the plane waves describing the exciton motion. The wave function $\psi(z_e,z_h,\rho,\theta)$ describes the relative motion of the electron and the hole in the exciton as well as their motion across the QW layer.

In Eq.~(\ref{eq6}), we used the cylindrical system of coordinates, in which the relative electron-hole motion in the xy plane is described by coordinates $\rho$ and $\theta$. Here $\theta$ is the angle between the vector $\overrightarrow{\rho}$ and the coordinate $x$ in the plane, and $|\vec{\rho}|$ is described by expression, 
%
\begin{equation}
\label{eq7}
\begin{aligned}
    \rho \equiv |\vec{\rho}|=\sqrt{(x_e-x_h)^2+(y_e-y_h)^2}.
\end{aligned}
\end{equation}

In what follows, we use the cylindrical symmetry of the exciton problem. In this case, we can further factorize the exciton wave function:
%
\begin{equation}
\label{eq8}
\begin{aligned}
    \psi(z_e,z_h,\rho,\theta)=\psi(z_e,z_h,\rho)e^{ik_\theta\theta},
\end{aligned}
\end{equation}
where $k_{\theta} = 0, 1, 2, … $ is the orbital quantum number describing the angular dependence of the exciton wave function~\cite{Khramtsov-JPh2016, Khramtsov-JAP2016}. Factorizations~(\ref{eq6}) and~(\ref{eq8}) allow one to simplify the SE for the exciton in a QW to a three-dimensional form~\cite{Khramtsov-JAP2016}:
%
\begin{equation}
\label{eq9}
\begin{aligned}
    &\left(-\frac{\hbar^2}{2\mu_{xy}}\left[\frac{\partial^2}{\partial\rho^2}-\frac{1}{\rho} \cdot \frac{\partial}{\partial\rho}+(1-{k_\theta}^2) \cdot \frac{1}{\rho^2}\right]-\right.\\
&\left.\frac{\hbar^2}{2m_e}\frac{\partial^2}{\partial z_e^2}-\frac{\hbar^2}{2m_{hz}}\frac{\partial^2}{\partial z_e^2}-\frac{q^2}{\varepsilon \cdot r}+V_e+V_h\right)\chi(\rho,z_e,z_h)\\
&=E_X \cdot \chi(\rho,z_e,z_h).
\end{aligned}
\end{equation}

Here $E_X$ is the exciton energy with respect to the bandgap energy $E_g$; $\mu_{xy}$ is the reduced effective mass of the exciton in the plane. In Eq.~(\ref{eq9}) we used substitution $\chi(z_e,z_h,\rho)=\psi(z_e,z_h,\rho) \cdot \rho$ to avoid divergence at the coinciding coordinates of the electron and the hole~\cite{Belov-Conf-2018, Belov-PRB2022, Khramtsov-JAP2016, Bataev-PRB2022, Chukeev-Semiconductors2024, Khramtsov-JPh2016}.

\vspace{2mm}
\section{Calculations of exciton characteristics in electric fields}
\vspace{-2mm}
\subsection{Exciton energy}
\vspace{-2mm}
The energy of an exciton is most often calculated using the variational method to solve the SE~\cite{Shinozuka-PRB1983, Andreani-PRB1990, Gerlach-PRB1998}. However, this method allows for the calculation of the exciton ground state predominantly. The calculations of excited states meet certain problems~\cite{Wilkes-NJPh2016, Wilkes-SM2017}. The accuracy of the calculated wave functions depends on the trial functions, which have to be chosen manually. Therefore, the accuracy is unknown a priori.

Recently, a numerical method of the SE solution has been extensively used, which does not require the trial functions~\cite{Khramtsov-JAP2016,Grigoryev-PRB2016,Grigoryev-SM2016,Belov-PhE2019, Bataev-PRB2022, Chukeev-PRB2024}. It is based on the finite difference method (FDM), which can provide accurate numerical results when the grid step is small enough. The central idea of the FDM is to represent the Hamiltonian of SE~(\ref{eq9}) as a matrix on the grid. To obtain the exciton state energies and wave functions, we must calculate the eigenvalues and eigenvectors of the matrix. In fact, only a few of the lowest exciton states that are observed in the experiment are of interest. They can be calculated using the well-known Arnoldi algorithm~\cite{Sorensen1995, Khramtsov-JAP2016}.

\subsection{Exciton binding energy}
\vspace{-2mm}
The obtained exciton energies allow one to also find the exciton binding energies. For this purpose, the energies of the quantum-confined electron, $E_e$, and hole, $E_h$, states should be calculated. The respective SE for a free electron in a QW is a one-dimensional one~\cite{Davies1997}:
%
\begin{equation}
\label{eq10}
\begin{aligned}
    \left(-\frac{\hbar^2}{2m_e}\frac{\partial^2}{\partial z_e^2}+V_e\right)\psi(z_e)=E_e \cdot \psi(z_e).
\end{aligned}
\end{equation}
A similar equation with replacing index $e \to h$ can be written for a free hole in the QW. The solution of Eq.~(\ref{eq10}) is easy and can be done with high accuracy~\cite{Davies1997}.

\subsection{Exciton-light coupling}
\vspace{-2mm}
The interaction of an exciton with light can be characterized by a constant $\Gamma_0$, which is the radiative decay rate of electromagnetic wave emitted by the exciton. In the experiment, a radiative broadening, $\hbar\Gamma_0$, of exciton resonances in reflection spectra can be measured~\cite{Grigoryev-SM2016, Bataev-PRB2022, Chukeev-PRB2024}. Therefore, the theoretical modeling of the broadening and its dependence on the electric field is of particular interest.

The exciton wave function obtained from the solution of the SE~(\ref{eq9}) allows one to calculate $\hbar\Gamma_0$ in the framework of the non-local dielectric response theory~\cite{Ivchenko2005}.
%
\begin{equation}
\label{eq11}
\begin{aligned}
    \hbar\Gamma_0=\frac{2\pi k}{\varepsilon}\left(\frac{e|p_{cv}|}{m_0\omega_0}\right)^2\left|\int_{-\infty}^{\infty}\Phi(z)e^{ikz}dz\right|^2.
\end{aligned}
\end{equation}
Here $k$ is the wave vector of the light wave, $\omega_0$ is the exciton resonance frequency, and $|p_{cv}|=m_0\cdot E_p/2$ is the matrix element of the momentum operator, calculated using the one-electron states of the conduction and valence bands, where $E_p = 28.8$ eV~\cite{Vurgaftman-JAP2001}; Function $\Phi(z)$ is the cross-section of the exciton wave function along the coinciding coordinates of the electron and the hole, $\Phi(z)=\psi(z_e=z_h=z,\rho=0)$. The integral in Eq. (\ref{eq11}) is frequently called an overlap integral of the light wave and the exciton wave function. This integral mainly controls the exciton-light coupling strength.

\subsection{Static dipole moment}
\vspace{-2mm}
In the electric field applied to the QW along the growth direction, the exciton is polarized. The electron and the hole in the exciton are shifted by the electric field to opposite potential barriers of the QW, thus creating a static dipole moment of the exciton. The dipole moment of exciton states can be calculated using the formula,
%
\begin{equation}
\label{eq12}
\begin{aligned}
    D_X=q\cdot \langle \psi(z_e,z_h,\rho) |(z_e-z_h)| \psi(z_e,z_h,\rho) \rangle .
\end{aligned}
\end{equation}

In the cylindrical coordinate system, this formula is transformed to
%
\begin{equation}
\label{eq13}
\begin{aligned}
    D_X=q \cdot \int |\psi(z_e,z_h,\rho)|^2(z_e-z_h)\cdot \rho d\rho d{z_e}d{z_h}.
\end{aligned}
\end{equation}

The dipole moment of an exciton is proportional to the average distance between the electron and the hole. We can expect that, in the general case, the dipole moment of an exciton in a QW nonlinearly depends on the applied electric field. Indeed, the distance cannot exceed the QW width. Therefore, the dipole moment magnitude should saturate at large fields.

\subsection{Exciton center-of-mass position}
\vspace{-2mm}
An exciton is a neutral quasiparticle, and its center of mass, at first glance, should be insensitive to an electric field. However, this is not the case. The fact is that the masses of the electron and the hole are different, which causes the exciton to shift towards a heavier particle. The calculation of the center-of-mass position is similar to that of the dipole moment,
%
\begin{equation}
\label{eq14}
\begin{aligned}
    Z=\left\langle \psi(z_e,z_h,\rho) \left| \frac{{m_e}{z_e}+{m_h}{z_h}}{m_e+m_h} \right| \psi(z_e,z_h,\rho)\right\rangle .
\end{aligned}
\end{equation}
As we demonstrate in the next section, the shift can be valuable, in particular, for the heavy-hole exciton.

\section{Results}
\vspace{-2mm}
We consider in this paper the heterostructure GaAs/Al$_x$Ga$_{1-x}$As with a parameter $x$ of $0.3$, representing the mass fraction of Al atoms in the barrier Al$_x$Ga$_{1-x}$As. The semiconductor layer GaAs can work as a QW for the electron and the hole because its bandgap energy, $E_g \approx 1.52$~eV~\cite{Vurgaftman-JAP2001,Bataev-PRB2022}, is smaller than that of the solid solution in the barriers. Moreover, the GaAs layer surrounded by the Al$_x$Ga$_{1-x}$As barriers created a potential well both for the electron and the hole, so that this is the type I heterostructure. In our calculations, we use the ratio of the potential well depths, $V_{e0}/V_{h0}=0.65/0.35$, commonly accepted in literature for the GaAs/Al$_x$Ga$_{1-x}$As heterostructures~\cite{Vurgaftman-JAP2001}.

Other parameters of the calculations are as follows: The effective mass of the electron is $m_e=0.067m_0$~\cite{Vurgaftman-JAP2001}. For the effective masses of the heavy and light holes, the following formulas must be used~\cite{Bataev-PRB2022}:
%
\begin{equation}
\label{eq15}
    \begin{aligned}
        m_{hhz}=\frac{m_0}{\gamma_1-2\gamma_2},~&m_{lhz}=\frac{m_0}{\gamma_1+2\gamma_2}\\
        m_{hhxy}=\frac{m_0}{\gamma_1+\gamma_2},~&m_{lhxy}=\frac{m_0}{\gamma_1-\gamma_2}
    \end{aligned}
\end{equation}
where, $\gamma_1=6.98$; $\gamma_2=2.06$  are the Luttinger parameters.~\cite{Vurgaftman-JAP2001}.

We consider single QWs of three different widths, $L = 30$, 50, and 100 nm. The Bohr radius of the exciton in the GaAs crystal is approximately equal to 15 nm~\cite{Ivchenko2005}. Therefore, the 30-nm and 50-nm QWs are both relatively narrow QWs, but these two widths of QWs give the excitons a certain activity space. The 100-nm QW is considerably wider than the exciton Bohr radius, so this QW gives much space for the exciton polarization in the electric field. Considering these three QWs, we demonstrate how the sensitivity of exciton states to the external electric field increases with the QW width increase.

\subsection{Stark shift and binding energy of exciton states in QW}
\begin{figure}[!h]
    \centering
    \includegraphics[scale=0.7]{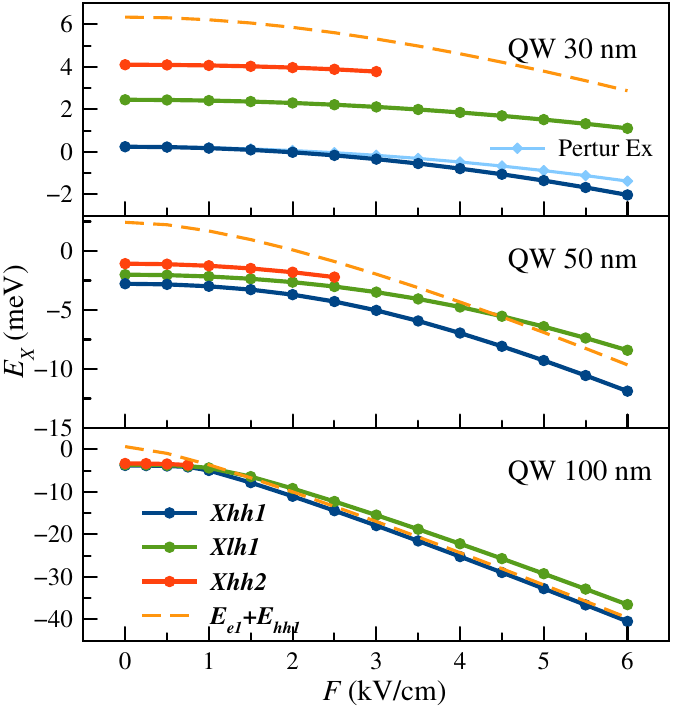} 
    \caption{The electric field dependence of exciton energies for different states in three QWs (solid lines with symbols). The dashed curves are the sum of the free electron and free heavy hole energies. The blue curve in the upper panel is the calculation in the framework of the perturbation theory with the Xhh1-Xhh2 coupling.}
    \label{fig:1}
\end{figure}

\vspace{-2mm}
The total energy of the exciton in the QW changes under the action of an external electric field. Results of macroscopic calculations of exciton energies in the studied QWs are shown in Fig.~\ref{fig:1}. Three exciton states, the first (Xhh1) and second (Xhh2) quantum-confined states of the heavy-hole exciton and the first quantum-confined state (Xlh1) of the light-hole exciton, can be reliably determined in the calculations. The higher-energy exciton states fall into the continuous spectrum of motion of free electrons and holes along the QW layer. Their numerical calculation by the method exploited in this work meets large problems described, e.g., in Refs~\cite{Bataev-PRB2022, Belov-PRB2022}. The low boundaries of the continuous spectrum for the pair (free electron + heavy hole, $E_{e1}+E_{hh1}$) are shown in Fig.~\ref{fig:1} by the orange dashed curves.

All exciton states experience a Stark shift to the low-energy region when an electric field is applied. The shift is relatively small for the 30-nm QW (units of meV), moderate for the 50-nm QW, and large for the 100-nm QW (tens of meV). Moreover, for the 30-nm QW, it can be well approximated by a parabolic dependence on the electric field in the whole range of electric field considered, $F \leq 6$ kV/cm. In the wider QWs, the parabolic fit is appropriate in the smaller field range. In particular, the Stark shift of the Xhh1 and Xlh1 exciton states tends to be linear in electric field, $F > 1.5$ kV/cm, for the 100-nm QW. The boundary of the continuous spectrum $E_{e1}+E_{hh1}$ also experiences the Stark shift so that the energy distance between the Xhh1 state and this boundary decreases. As a result, the Xhh2 state falls into the region of the continuous spectrum and cannot be reliably determined above some values of the electric field, as indicated in Fig.~\ref{fig:1}.

When the electric field strength is small enough, the change of the exciton energy, the Stark shift, can be described in the framework of the perturbation theory~\cite{Landau-Lifshits}. The Stark shift is determined by the potential energy operator, $qF(z_e-z_h)$, and is described by the second-order perturbation expression:
%
\begin{equation}
\label{eq16}
\begin{aligned}
    E_{x1}&=E_{x1}^0-\\&(qF)^2~\cdot\sum_{n\neq1}\frac{\left| \left\langle \psi(z_e,z_h,\rho)_n^0|(z_e-z_h)|\psi(z_e,z_h,\rho)_1^0 \right\rangle \right|^2}{E_{xn}^0-E_{x1}^0}\\
    &\equiv E_{x1}^0-AF^2.
\end{aligned}
\end{equation}

Here $E_{x1}^0$ and $E_{xn}^0$ are the ground state and n-th excited state energies of the exciton in the QW without external perturbation. Similarly, $\psi(z_e,z_h,\rho)_1^0$ and $\psi(z_e,z_h,\rho)_n^0$ are respective wave functions in the cylindrical coordinate system in the absence of the electric field. The summation in Eq.~(\ref{eq16}) is performed over all the excited states. It can be seen from this expression that, under the condition of satisfying the perturbation theory, the Stark energy shift should be a quadratic function of the external electric field strength.

It is instructive to estimate the magnitude of the Stark shift of the Xhh1 state, taking into account only one term of the sum in Eq.~(\ref{eq16}), namely, the coupling the Xhh1 and Xhh2 states. The curve calculated in this way for the 30-nm QW is shown in the upper panel of Fig.~\ref{fig:1} by the blue color. As seen, this curve shows a bit smaller Stark shift than that in the exact numerical calculations. The curvature coefficient $A$ [see Eq.~(\ref{eq16})] obtained in the perturbation theory is $A = -0.045$ meV/(kV/cm)$^2$. This value is about $70\%$ of that found from exact numerical calculations. This means that the Xhh1-Xhh2 coupling gives the main contribution to the Stark shift of the Xhh1 exciton state in this QW. In the 50-nm and 100-nm QWs, the perturbation theory is applicable in the smaller electric field range, and the Xhh1-Xhh2 coupling describes, respectively, $47\%$ and $7\%$ of the Stark shift curvature coefficient.

The exciton energy can be described as the sum of the free electron and free hole energies minus their binding energy,
%
\begin{equation}
\label{eq17}
    \begin{aligned}
        E_X=E_e+E_h-E_b.
    \end{aligned}
\end{equation}

\begin{figure}[!h]
    \centering
    \includegraphics[width=0.92\linewidth]{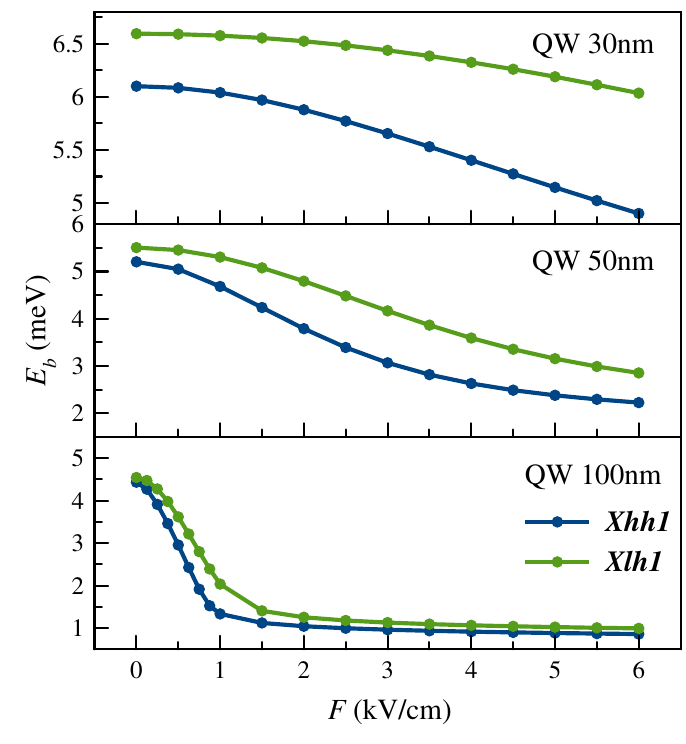}
    \caption{Electric field dependence of the binding energy of exciton states in three QWs.}
    \label{fig:2}
\end{figure}

When the exciton is stretched by the electric field, the distance between the electron and hole that constitute it gradually increases, causing the binding energy to gradually approach zero~\cite{Belov-PhE2019,Belov-PRB2022}. As it is seen in Fig.~\ref{fig:1}, the energy distance between the $E_e+E_{hh1}$ and $E_X$ of Xhh1 curves gradually decreases with the electric field increase, which indicates the decrease of the binding energy. The dependencies of the binding energies of the Xhh1 and Xlh1 exciton states on the electric field are shown in Fig.~\ref{fig:2} for all three QWs studied.

The decrease of the binding energy is different in different QWs. It is small for the 30-nm QW and much larger for the wider QWs. Moreover, the binding energy decreases down to a finite value, rather than to zero, in the 100-nm QW. This phenomenon is explained by the fact that the electron and the hole, which are moved apart from each other by the applied electric field, are blocked by the QW boundaries. Correspondingly, their Coulomb interaction remains finite. The Coulomb energy,
%
\begin{equation}
\label{eq18}
    \begin{aligned}
        E_C=\left\langle \psi(z_e,z_h,\rho)\left |\frac{q^2}{\varepsilon \cdot r}\right |\psi(z_e,z_h,\rho) \right\rangle ,
    \end{aligned}
\end{equation}
obtained using the calculated wave functions for the Xhh1 and Xlh1 exciton states in the 100-nm QW in an electric field $F = 6$ kV/cm are, respectively, 1.27 meV and 1.37 meV. The low limit of the Coulomb energy for an electron and a hole as point charges at a distance $L = 100$ nm, $E_C = q^2/(\varepsilon L) = 0.98$ meV. Here $\varepsilon = 12.53$ is the dielectric constant of GaAs~\cite{Gerlach-PRB1998}. The binding energy obtained in the exact numerical calculations for the 100-nm QW in a strong electric field, $F = 6$~kV/cm, is a bit smaller: $E_b = 0.86$~meV for the Xhh1 exciton state and $E_b = 0.99$ meV for the Xlh1 exciton state. This seems contradictory. The result is explained by the contribution of kinetic energy, $E_{kin}$, of the charged carriers to the binding energy: $E_b = |- E_C + E_{kin}|$. Comparing these values, we can conclude that the kinetic energy has little effect on the binding energy of excitons in strong electric fields.

\begin{figure}[!h]
    \centering
    \includegraphics[width=1\linewidth]{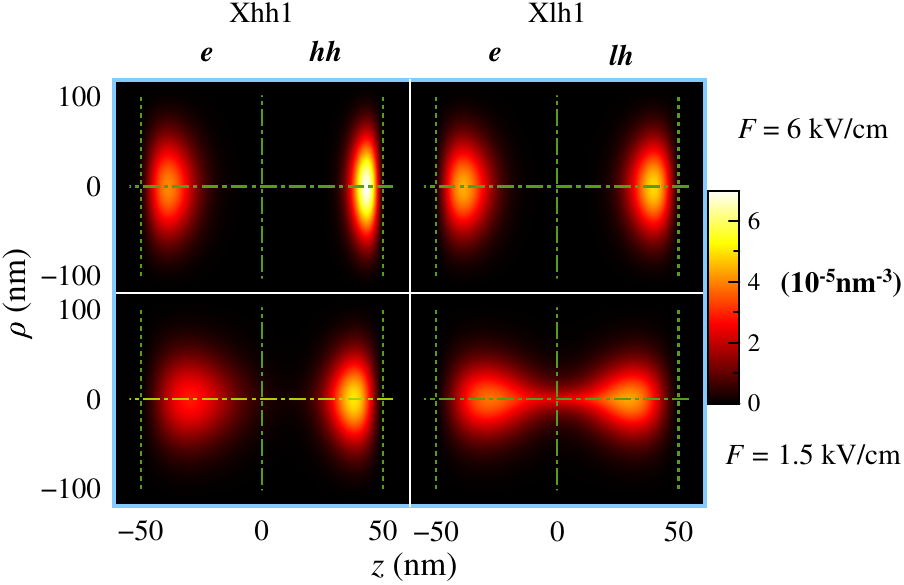}
    \caption{The sum of the electron and hole density distributions for the Xhh1 and Xlh1 exciton states, $|\psi_e(z_e,\rho)|^2+|\psi_h(z_h,\rho)|^2$, in the 100-nm QW for the case of electric field $F = 1.5$ kV/cm (lower panels) and $F = 6$ kV/cm (upper panels). The calculation step is 1 nm.}
    \label{fig:3}
\end{figure}

The effect of the QW boundaries is illustrated in Fig.~\ref{fig:3} for the 100-nm QW in more detail. The figure shows the charged carrier density distributions. They are calculated by integrating the exciton wave function squared over the coordinate z of another particle~\cite{comment419}. E.g., the expression for electron density distribution reads,
%
\begin{equation}
\label{eq19}
    \begin{aligned}
    \rho_e=|\psi_e(z_e,\rho)|^2=\int|\psi(z_e,z_h,\rho)|^2dz_h.
    \end{aligned}
\end{equation}
A similar expression for the hole density distribution $\rho_h$ is obtained by replacing indexes $e \leftrightarrow h$. It is worth noting that the exciton wave functions are normalized, $\langle\psi(z_e,z_h,\rho)|\psi(z_e,z_h,\rho)\rangle$, so that the color in Fig.~\ref{fig:3} reflects real density in units of nm$^{-3}$. As seen, when the applied electric field increases above 1.5 kV/cm, the electron and hole densities are concentrated near the QW boundaries and blocked by them. Correspondingly, the relative distance between the two charged carriers remains almost unchanged at large electric fields.

Fig.~\ref{fig:3} also shows that the density distribution maximum of the heavy hole in the Xhh1 exciton state is closer to the QW boundary than that of the light hole in the Xlh1 exciton state. This is because the quantization energy of a particle is inversely proportional to its effective mass. Correspondingly, the energy of the heavy hole is smaller than that of the light hole. The electric field transforms the original rectangular QW into a triangular QW~\cite{Miller-PRB1985}. Therefore, the wave function of the light holes with greater energy can be more widely distributed in the triangle QW. Similarly, the electron density is more widely distributed because the electron mass is smaller than the masses of the light and heavy holes.

\subsection{Effect of the external electric field on the exciton-light coupling}
\begin{figure}
    \centering
    \includegraphics[width=1\linewidth]{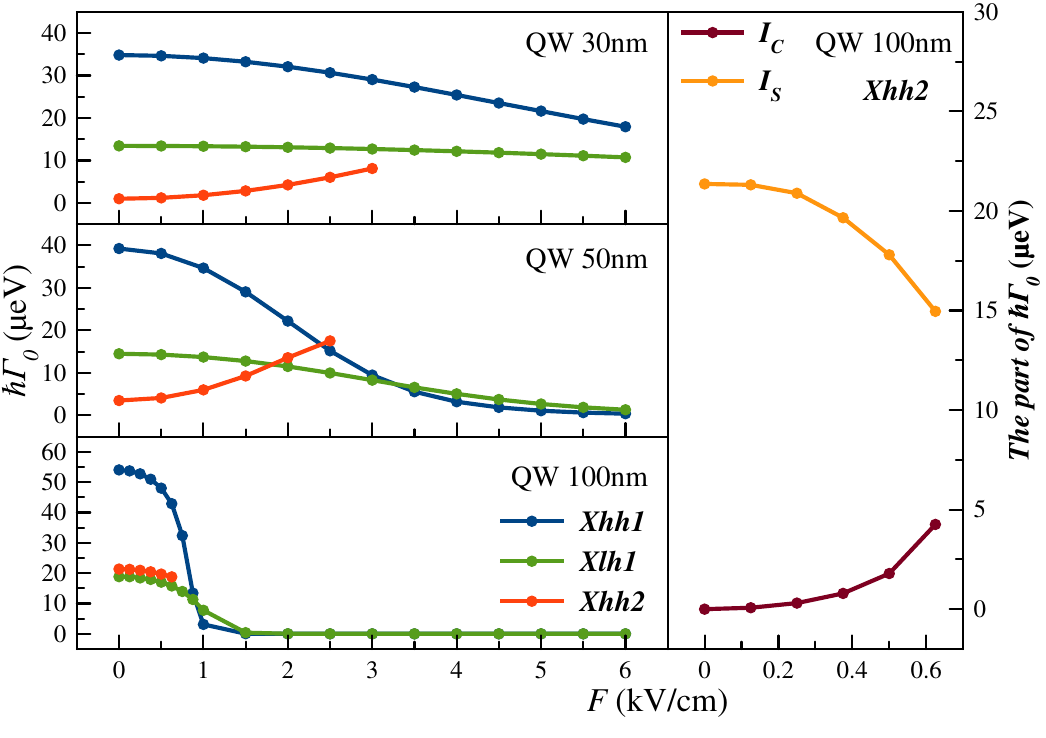}
    \caption{Dependence of radiative broadening, $\hbar\Gamma_0$, on the applied electric field. The right panel shows different contributions to the exciton-light coupling for the Xhh2 state in the 100-nm QW.}
    \label{fig:4}
\end{figure}

\vspace{-2mm}
The ability of the exciton states (Xhh1, Xlh1, Xhh2) in a QW to couple with incident photons changes when the states are stretched by an applied electric field. We have modeled the influence of the electric field on the exciton-light coupling constant, $\hbar\Gamma_0$. Results are shown in Fig.~\ref{fig:4}. We have to note that the $\hbar\Gamma_0$ of the Xhh2 exciton state obtained in our calculations is unreliable when the external electric field F is greater than 3 kV/cm, 2.5 kV/cm, and 0.75 kV/cm in the 30-nm, 50-nm, and 100-nm QWs, respectively. This is due to the overlap of this state with the continuum of states of the uncoupled electron-hole pairs.

As seen in Fig.~\ref{fig:4}, the radiative broadening $\hbar\Gamma_0$ of the Xhh1 and Xlh1 exciton states decreases as the external electric field increases. This effect is most pronounced for the 100-nm QW. The exciton-light coupling drops for the Xhh1 state from $\hbar\Gamma_0 = 54$~$\mu$eV down to $0.04$~$\mu$eV and for the Xlh1 state from 19~$\mu$eV down to 0.4~$\mu$eV even in a small electric field of 1.5 kV/cm. The fast drop of the exciton-light coupling is explained by the fact that the average distance between the electron and hole inside the exciton state gradually increases with the electric field rise, see Fig.~\ref{fig:3}. Accordingly, the probability of the electron and the hole being located at the same point in space, which determines the exciton-light coupling [see Eq. (\ref{eq11})], decreases. However, the behavior of $\hbar\Gamma_0$ for the Xhh2 exciton state is different. Namely, $\hbar\Gamma_0$ gradually increases for the 30-nm and 50-nm QWs as the external electric field increases. For the 100-nm QW, it remains almost constant in the electric fields studied.

The theoretically obtained phenomena are consistent with those experimentally observed in Ref.~\cite{Chukeev-PRB2024} for radiative broadening of exciton states in a 30-nm QW in external electric fields. According to this work, the radiative broadening of the Xhh1 and Xlh1 exciton states decreases with the increase of the external electric field because the overlap of the electron and hole wave functions weakens with the increase of the external electric field. This is consistent with the results shown in Fig.~\ref{fig:3}. At the same time, the experiments for the 30-nm QW show~\cite{Chukeev-PRB2024} that the radiative broadening of the Xhh2 exciton state increases with the increase of the external electric field. Our calculations for the 30-nm QW are consistent with the experimental results.

\begin{figure}
    \centering
    \includegraphics[width=0.9\linewidth]{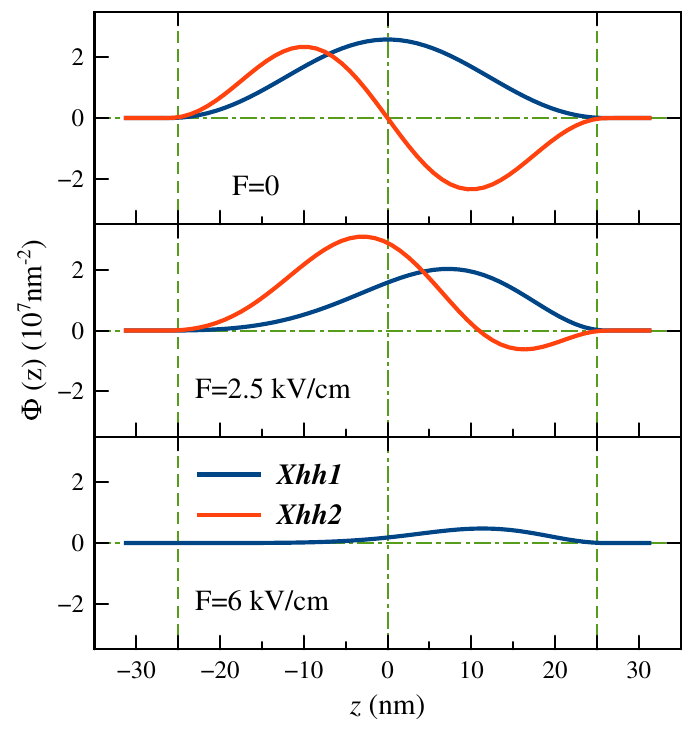}
    \caption{The cross section $\Phi(z)$ of the wave function of the exciton state Xhh2 in the 50-nm QW in different electric fields indicated in the panels. The calculation step is 1 nm.}
    \label{fig:5}
\end{figure}

Let us analyze the theoretically predicted behavior of exciton-light coupling more quantitatively. As it is shown in Eq.~(\ref{eq11}), the main parameter that determines the exciton-light coupling is the overlap of the exciton function $\Phi(z)$ and $e^{iqz}$ describing the light wave. According to Euler's formulas and the definition of the module of a complex number, the integral squared can be expressed as:
%
\begin{equation}
\label{eq20}
    \begin{aligned}
        &\left | \int_{-\infty}^{\infty}\phi(z)e^{iqz}dz \right |^2\equiv I_C+I_S\\
        &=\left | \int_{-\infty}^{\infty} \Phi(z) \cdot \cos(qz)dz \right |^2+\left | \int_{-\infty}^{\infty} \Phi(z) \cdot \sin(qz)dz \right |^2.
    \end{aligned}
\end{equation}

Functions $\Phi(z)$ for the Xhh1 and Xlh1 exciton states are symmetric relative to the QW center in zero electric field. Accordingly, only the first term in Eq.~(\ref{eq20}), $I_C$, contributes to $\hbar\Gamma_0$. When the electric field is applied, functions $\Phi(z)$ are distorted, and both terms, Ic and Is, can contribute to the exciton-light coupling. Examples of functions $\Phi(z)$ of the Xhh1 state in different electric fields for the 50-nm QW are shown in Fig.~\ref{fig:5}. However, the amplitude of $\Phi(z)$ considerably drops in strong electric fields (see an example in the low panel of Fig.~\ref{fig:5}). This is due to the decrease in overlap of the electron and hole density distributions, as shown in Fig.~\ref{fig:3}. This effect results in strong drops of $\hbar\Gamma_0$.

The behavior of the exciton-light coupling of the Xhh2 state shown in Fig.~\ref{fig:4} can be understood considering again Eq.(\ref{eq11}) and the profile of the function for this state. At zero electric field, the function is anti-symmetric relative to the QW center, see Fig.~\ref{fig:5}. Correspondingly, only the second term, $I_S$, of Eq.~(\ref{eq20}) contributes to the $\hbar\Gamma_0$. This contribution is relatively small for the 30-nm and 50-nm QWs due to the small QW width relative to the light wavelength. In the 100-nm QW, the contribution is considerably greater due to the larger overlap of $\Phi(z)$ and the light wave. When the electric field is applied, the function $\Phi(z)$ is modified as shown in Fig.~\ref{fig:5}. Accordingly, the overlap integral $I_C$ starts to grow, which results in an overall increase of the exciton-light coupling in the 30-nm and 50-nm QWs. However, in the case of the 100-nm QW, the increase of integral $I_C$ is compensated by the rapid decrease of integral $I_S$ as shown in Fig.~\ref{fig:4}. As a result, the $\hbar\Gamma_0$ is almost unchanged in the electric field range studied.

\subsection{Exciton dipole moment in external electric field}
\begin{figure}[!h]

    \centering
    \includegraphics[width=0.9\linewidth]{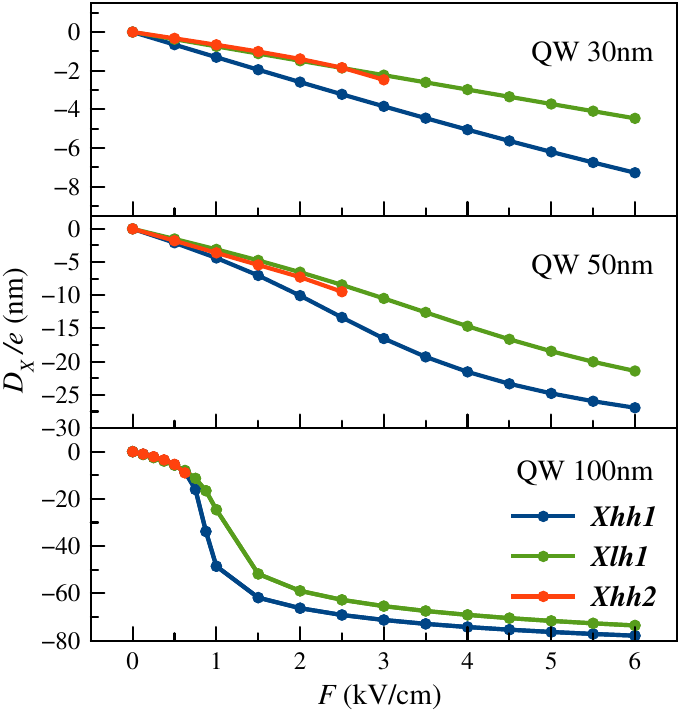}
    \caption{Dependence of the exciton dipole moment on the applied electric field.}
    \label{fig6}
\end{figure}

\vspace{-2mm}
As mentioned above, the static dipole moment of the exciton state is proportional to the average distance between the electron and the hole inside the exciton. We have studied the relationship between the static dipole moment of the exciton state and the external electric field. In the individual QWs considered in this work, the absolute values of the static dipole moments of all three exciton states gradually increase in absolute value from 0 with the increase of the electric field, see Fig.~\ref{fig6}. The negative values in Fig.~\ref{fig6} only indicate the direction of the dipole moment. The smaller the QW width is, the easier the exciton is blocked by the boundaries of the QW. Therefore, the dipole moments of all the exciton states increase linearly and slowly in the 30-nm QW. The maximum value of $D_X/q$, the dipole moment in units of the electron charge, is only -7 nm for the Xhh1 state and about -4 nm for the Xlh1 state, which is much smaller than the exciton Bohr radius ($a_B \approx 15$ nm).

When the width of the QW increases to 50 nm, the exciton state becomes more sensitive to the external electric field. For the electric field $F \geq 4$ kV/cm, the dipole moment of the Xhh1 exciton state begins to change nonlinearly with the increase of the electric field. This means that the electron and hole in the exciton state are stretched and blocked by the QW boundaries. When the width of the QW reaches 100 nm, the QW becomes a wide QW for the exciton. Even in a tiny electric field, $F = 1.5$ kV/cm, the electron and the hole in the Xhh1 and Xlh1 exciton states move close to the QW boundaries. The average distance between the electron and hole becomes large, $|D_X/q| > 50$ nm, which is larger than the exciton Bohr radius. The electron and hole charges are effectively separated, as shown in Fig.~\ref{fig:3}. Further strong increase of the dipole moment is blocked by the QW boundaries. This behavior of the dipole moment is consistent with that of the exciton energy $E_X$, the exciton binding energy $E_b$, and the radiative broadening shown in Figs.~\ref{fig:1},~\ref{fig:2}, and~\ref{fig:4}, respectively.

\subsection{Shift of exciton center of mass in electric field}
\begin{figure}[h!]
    \centering
    \includegraphics[width=0.9\linewidth]{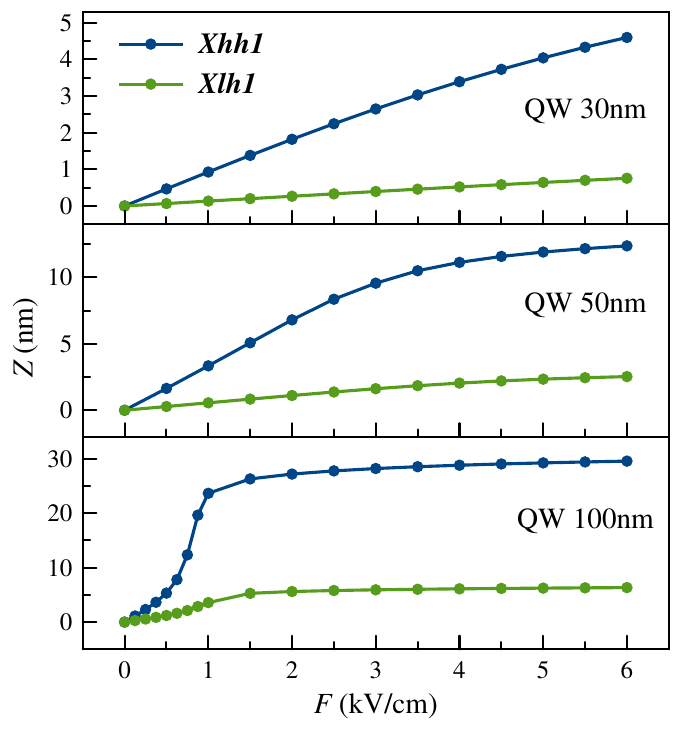}
    \caption{\label{fig7}Electric field dependence of the center of mass position of exciton states in separate QWs of different widths.}
\end{figure}

\vspace{-2mm}
Due to the different effective masses of an electron and a hole, the application of electric fields leads to the movement of the center of mass of excitons. This effect is described by Eq.~(\ref{eq14}). It is obvious from this expression that there is no shift of the center of mass when $m_e = m_h$ because the electron and the hole are moving in opposite directions so that
%
\begin{equation}
\label{eq21}
    \begin{aligned}
        Z=\left \langle \psi(z_e,z_h,\rho) \left | \frac{z_e+z_h}{2} \right | \psi(z_e,z_h,\rho) \right \rangle=0,
    \end{aligned}
\end{equation}
for all non-zero electric fields. The greater the difference in mass of charged carriers, the greater the shift $Z$ should be. Here we have calculated the center of mass position for the Xhh1 and Xlh1 exciton states in all three QWs when the external electric field is applied. Results are shown in Fig.~\ref{fig7}.

As seen in the figure, the shift $Z$ of the heavy-hole and light-hole excitons considerably differs in all three QWs. In the 30-nm QW, it is almost linear in the electric fields studied and relatively small compared to the QW width. At the same time, the Xhh1 exciton shift is approximately 6 times greater than the Xlh1 exciton shift. For the case of the 50-nm QW, the shift of the Xhh1 exciton becomes sublinear at $F > 3$ kV/cm, while the Xlh1 exciton position almost linearly depends on the electric field. The ratio of the exciton shifts, $R = Z_{Xhh1}/Z_{Xlh1} = 4.9$, for $F = 6$ kV/cm. In the case of the 100-nm QW, the shifts of both excitons slowly increase at electric fields $F > 1.5$ kV/cm, and their ratio, $R = 4.7$, at $F = 6$ kV/cm. We should note that the ratio of the Xhh1 and Xlh1 exciton masses is noticeably smaller: $R_m = (m_e + m_{hh})/(m_e+m_{lh}) = 2.66$. Thus, a strong shift of the heavy-hole exciton compared to that of the light-hole exciton is associated not only with the difference in their masses but also with a stronger localization of the wave function of the heavy-hole exciton.

The observed nonlinear behavior of the exciton center of mass shift is consistent with the behavior of other exciton characteristics shown in Figs.~\ref{fig:2},~\ref{fig:4}, and~\ref{fig6}. It is instructive again to evaluate the maximum shift of the exciton center of mass considering the electron and hole as point charges. In strong electric fields, they are located near the QW boundaries, that is, $z_e = -50$ nm and $z_h = 50$ nm in the 100-nm QW. Simple calculations show that the center of mass positions of the Xhh1 and Xlh1 exciton states are, respectively, 34 nm and 7.3 nm. These values are close to those obtained in the exact numerical calculations, $Z_{Xhh1} = 29.5$ nm and $Z_{Xlh1} = 6.32$ nm, for the 100-nm QW.

\section{Modeling of reflection spectra}

The properties of exciton states in QW in heterostructures can be studied in experiments through their reflection spectra~\cite{Chukeev-PRB2024}. The exciton characteristics numerically obtained in previous sections allow us to model reflection spectra of heterostructures containing QWs with 30-nm, 50-nm, and 100-nm widths, which could be observed experimentally.

According to the nonlocal dielectric response theory~\cite{Ivchenko2005}, the intensity reflection coefficient, $R$, of a QW-heterostructure can be expressed as follows:
\begin{equation}
\label{eq_Reflectance}   
R(\hbar\omega)=\left|\frac{r_s+r_{QW}(\hbar\omega)}{1+r_s\cdot r_{QW}(\hbar\omega)}\right|^2.
\end{equation}
Here, $r_s$ is the amplitude reflection coefficient from the surface of the heterostructure sample, $r_{QW}$ is the amplitude reflection coefficient of the QW, and $\hbar\omega$ is the photon energy of incident light.

The coefficient $r_s$ weakly depends on the photon energy and can be fixed for the GaAs surface in the spectral range we consider~\cite{Grigoryev-SM2016}, $r_s\approx -0.565$. The coefficient $r_{QW}$ is a function of the incident photon $\hbar\omega$, which can be obtained using the radiative broadening $\hbar\Gamma_{0j}$ of the exciton resonance $j$, the non-radiative broadening $\hbar\Gamma_j$, the phase $\phi_j$, and the exciton energy $E_{Xj}$:
\begin{equation}
\label{eq_r_QW}   
r_{QW}=\sum_j\frac{i\hbar\Gamma_{0j}e^{i2\phi_j}}
{(E_g+E_{Xj})-\hbar\omega-i\hbar(\Gamma_j+\Gamma_{0j})}.
\end{equation}
Here, $E_g = 1519.4$~meV, the same as that in the Bataev's paper~\cite{Bataev-PRB2022}, index $j = 1,$~2,~3,... numerates exciton states. Parameter $\phi_j$ is the exciton resonance phase~\cite{Grigoryev-SM2016, Chukeev-PRB2024}, which is given by the expression:
\begin{equation}
\label{eq_exciton_resonance_phase}   
\tan(\phi_j)=\frac{\int \Phi_j(z) \sin(qz) dz}{\int \Phi_j(z) \cos(qz) dz}.
\end{equation}
Besides phase $\phi_j$, which is determined by the symmetry of the exciton wave function, another phase, $\phi_0$, can modify the profile of exciton resonances~\cite{Shapochkin-PRA2019}. This phase shift is caused by the propagation of the light wave from the structure surface to the QW. In the theoretical calculation, we stipulate that the phase $\phi_0$ is 0 for the purpose of simplifying the problem.

Parameter $\hbar\Gamma_j$ in Eq.~(\ref{eq_r_QW}) is the nonradiative broadening of the exciton resonance $j$. This parameter phenomenologically describes the contribution of all mechanisms of homogeneous broadening of the exciton resonance~\cite{Kurdyubov-PRB2021}, which is always observed in experiments. The value of $\hbar\Gamma_j$ cannot be obtained from the solution of the Schrödinger equation for excitons. Therefore we choose its value manually, $\hbar\Gamma_j = 120$~$\mu$eV for all resonances in the 30-nm and 50-nm QWs and 60~$\mu$eV for the resonances in the 100-nm QW. These values of the nonradiative broadening are typical for the high-quality GaAs/AlGaAs heterostructures with relatively wide QWs~\cite{Trifonov-PRB2015, Trifonov-PRL2019, Shapochkin-PRA2019, Bataev-PRB2022}.

\begin{figure*}
\includegraphics[width=1.0\linewidth]{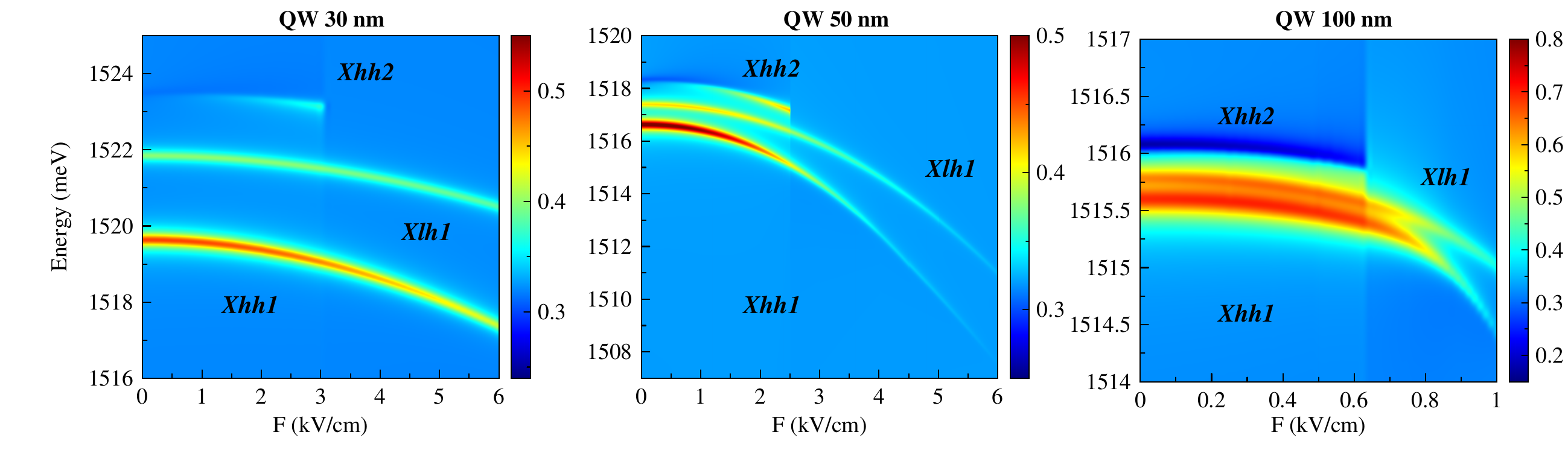}
\caption{\label{fig:8}Reflection spectra modeled for the QW widths of 30, 50, and 100 nm. The exciton states Xhh1, Xlh1, and Xhh2 are taken into account.}
\end{figure*}

To model the reflection spectra, we used numerical results for exciton energies $E_X$ and radiative broadening $\hbar\Gamma_0$, shown in Figs.~\ref{fig:1} and \ref{fig:4}, respectively. The discrete numerical data were interpolated to obtain smooth dependencies on the electric field. For the Xhh2 exciton states, we used only the data reliably obtained in the calculations in the limited ranges of the electric field, $F<3$ kV/cm for the 30-nm QW, $F<2.5$ kV/cm for the 50-nm QW, and $F<0.625$ kV/cm for the 100-nm QW.

The modeled reflection spectra are shown in Fig.~\ref{fig:8}. The light blue color in the figure represents the background reflection from the heterostructure surface. For the GaAs surface, its value, $R_s = |r_s|^2 \approx 0.32$. The exciton states in reflection spectra are represented as resonant features. These resonances are depicted in the green to red colors when they have a peak-like shape with an increased reflection coefficient and in dark blue for the dip in the reflection.

The Xhh1 and Xlh1 exciton resonances are represented as peaks of reflection. Their full width at half maximum (FWHM), $\Delta E$, is determined by the radiative $\hbar\Gamma_0$ and non-radiative $\hbar\Gamma$ broadenings of the exciton resonance, as well as by the amplitude reflection $|r_s|$,~
$\Delta E = 2\hbar[(\Gamma_0+\Gamma)+|r_s|\Gamma_0]$~\cite{Shapochkin-PRA2019}. The Stark shift of these resonances reproduces that shown in Fig~\ref{fig:1}. However, their visibility rapidly drops with the electric field increase, in particular, for the 50-nm and 100-nm QWs. This is due to the electric-field-induced decrease of the exciton-light coupling strength shown in Fig.~\ref{fig:4}. These resonances become almost invisible for the 100-nm QW at electric field $F > 1$~kV/cm; therefore, this field range is not shown in the right panel of Fig.~\ref{fig:8}. Visibility of these resonances in experiments can drop even faster because of an additional nonradiative broadening of the resonances with the increase of the electric field~\cite{Chukeev-PRB2024}. The Xhh and Xlh resonance can overlap in the 100-nm QW due to the nonradiative broadening.

The Xhh2 exciton state in the studied QWs is seen differently in the reflection spectra compared to the Xhh1 and Xlh1 states. At small electric fields, the Xhh2 state is observed as a dip of reflection for all three QWs, see~Fig.~\ref{fig:8}. For example, the reflection coefficient for a heterostructure with the 100-nm QW at the Xhh2 exciton resonance is about 0.18 at $F=0$~kV/cm. When the external electric field gradually increases, the Xhh2 resonance is transformed to a peak. The maximum amplitude of the Xhh2 resonant reflection reaches 0.42 for the 50-nm QW in the electric field of 2.5~kV/cm. Possibly, it can increase further in the larger electric fields, but we have not been able to reliably calculate the characteristics of the Xhh2 exciton for such electric fields. In the heterostructure with the 100-nm QW, the Xhh2 resonance is observed as a dip in the whole range of electric fields we managed to study.

The general shape of the exciton resonance reflection for a heterostructure with the 50-nm QW is similar to that for the 30-nm QW, which indicates a gradual change in the characteristics of exciton states from characteristics in a relatively narrow QW to characteristics in a wide QW.

\section{Conclusion}
\vspace{-2mm}
We used the fourth-order finite difference method to accurately solve the Schr{\"o}dinger equation that describes the quantum mechanical behavior of exciton states (Xhh1, Xlh1, and Xhh2) in the GaAs/Al$_x$Ga$_{1-x}$As ($x=0.3$) QWs with widths of 30, 50, and 100 nm in an external electric field applied along the structure axis. The Stark shift of exciton states is found to gradually increase from units of meV in the 30-nm QW to tens of meV in the 100-nm QW at $F = 6$ kV/cm. Under small electric fields, the Stark shift can be described by perturbation theory. The Xhh2 state was studied in a smaller range of electric fields. At large electric fields, for example, at $F > 3$ kV/cm for the 30-nm QW, it overlaps with a continuous spectrum of states of the free electrons and holes and cannot be found by the numerical algorithm used.

The binding energy of the Xhh1 and Xlh1 gradually decreases as the external electric field increases from 0 to 6 kV/cm. This decrease is greater the wider the QW. For the 100-nm QW, an abrupt, threshold-like, decrease in the binding energy is observed in the electric fields of $0.5 \to 1$ kV/cm. This behavior indicates a break in the Coulomb coupling of the electron and the hole in the exciton, i.e., dissociation of the exciton. At the same time, the binding energy decreases down to a finite, rather than zero, value of about 1 meV. This is due to the blocking effect of the QW boundaries when the average distance between the electron and the hole inside the exciton cannot exceed the width of the QW at most.

The applied electric field polarizes the excitons. The static dipole moment of the Xhh1 and Xlh1 excitons linearly rises (in absolute value) in the 30-nm QW and abruptly increases at $F \approx 1$ kV/cm in the 100-nm QW. Then the dipole moment weakly changes at larger electric fields in this QW, which is also due to the blocking effect of the QW boundaries. 

The exciton-light coupling constant $\hbar\Gamma_0$ of the Xhh1 and Xlh1 states slowly decreases in the narrow QW and abruptly decreases down to zero at $F \textgreater 1$ kV/cm in the 100-nm QW. This indicates a strong decrease in the overlap of the electron and hole densities in the excitons induced by the exciton polarization. The exciton-light coupling constant of the Xhh2 state is small for the narrow QW and comparable with that of the Xlh1 state for the 100-nm QW in the absence of the electric field. This constant increases with the electric field increase in the 30-nm and 50-nm QWs and remains almost unchangeable in the 100-nm. This behavior is explained by peculiarities of the overlap of the sine-like and cosine-like parts of the light wave with the Xhh2 exciton state, whose wave function is anti-symmetric at zero field and is strongly modified when the electric field is applied.

The applied electric field shifts the center of gravity of the excitons, although the exciton is a neutral quasiparticle. This effect is caused by a difference in the masses of the electron and the hole in the exciton. The heavy hole/electron mass ratio $m_{hh}/m_e = 5.2$, that is several times larger than that for the light hole/electron ($m_{lh}/m_e = 1.34$). Therefore, the shift of the center of mass of a heavy-hole exciton with an increase in the field occurs several times stronger for all three QWs. For a wide QW (100 nm), the threshold behavior of the effect is again observed in the field $F = 1$ kV/cm, above which the displacement of the center of mass slows down. This is also explained by the limitation of the movement of the electron and the hole in the QW.

The obtained results allowed us to model reflection spectra, which can be observed experimentally. The modeling shows that the visibility of exciton resonances gradually decreases with the electric field increase, in particular, for wide QWs. For the 100-nm QWs, the resonances are hardly visible at electric fields $F > 1$~kV/cm.

\vspace{3mm}
\begin{acknowledgments}
\vspace{-2mm}
Financial support from the Russian Science Foundation, grant No. 19-72-20039, and the Saint-Petersburg State University, grant No. 122040800257-5, is acknowledged. Research was carried out using computational resources provided by Resource Center "Computer Center of SPbU" (http://www.cc.spbu.ru/en). Shiming Zheng would like to thank the support from China Scholarship Council. The authors thank M. A. Chukeev and D. K. Loginov for fruitful discussions and Alexander Levantovsky for providing the advanced plotting and curve fitting program MagicPlot.
\end{acknowledgments}

\nocite{*}

\bibliography{Shiming}

\begin{thebibliography}{44}
\bibitem [{com()}]{comment419}%
  \BibitemOpen
  \href@noop {} {}\bibinfo {note} {Because of different masses of electrons and holes, the distribution along the $\rho$ coordinate should be different for different particles. We do not consider here this effect because we are mainly interested in the distribution along the $z$ coordinate.}\BibitemShut {Stop}%
\end{thebibliography}

\end{document}